\documentclass[aip,amsmath,amssymb,reprint]{revtex4-2}

\usepackage{graphicx}
\usepackage{dcolumn}
\usepackage{bm}
\usepackage{color}

\usepackage[utf8]{inputenc}
\usepackage[T1]{fontenc}
\usepackage{mathptmx}
\usepackage{etoolbox}

\begin{document}

\preprint{AIP/123-QED}

\title{Reciprocal transverse Seebeck and Peltier effects in MoSi$_2$}

\author{Aina~Sakamoto}
\author{Shoya Ohsumi}
\author{Shogo~Yoshida}
\email{s\_yoshida@rs.tus.ac.jp}
\author{Ryuji~Okazaki}
\email{okazaki@rs.tus.ac.jp}
\affiliation{Department of Physics and Astronomy, Tokyo University of Science, Noda 278-8510, Japan}

\author{Yoshiki~J.~Sato}
\affiliation{Graduate School of Science and Engineering, Saitama University, Saitama 338-8570, Japan}

\date{\today}

\begin{abstract}

Transverse thermoelectric conversion, 
in which charge and heat currents are interconverted in orthogonal directions, 
has recently attracted increasing attention 
as an alternative to conventional longitudinal thermoelectric conversion. 
Although transverse Seebeck effect has been investigated in various materials, 
its reciprocal process, transverse Peltier effect,
remains much less explored, 
and direct comparison of the two effects in the same homogeneous material is still limited. 
Here, we experimentally investigate both transverse Seebeck and Peltier effects in a single crystal of MoSi$_2$,
an axis-dependent conduction polarity material
in which the sign of the Seebeck coefficient varies depending on the crystallographic axis.
Multi-terminal voltage measurement determines the direction of the induced internal electric field and establishes
the transverse Seebeck response. 
We also directly visualize 
the transverse temperature modulation consistent with
the corresponding transverse Peltier effect
by using lock-in thermography under an applied ac current.
These results establish reciprocal transverse thermoelectric conversion in a homogeneous nonmagnetic single crystal without engineered interfaces.

\end{abstract}

\maketitle


Thermoelectric transport is governed by reciprocal conversion between charge
and heat currents. In the linear-response regime, 
the heat-current density $\bm J_{\rm q}$ can
be written as
\begin{align}
 \bm J_{\rm{q}}
 =\boldsymbol \Pi \bm J_{\rm{c}}
 -\boldsymbol \kappa \nabla T ,
\end{align}
where
$\bm J_{\rm{c}}$ is the charge-current density,
$\boldsymbol\Pi$ is the Peltier tensor, 
$\boldsymbol \kappa$ is the thermal-conductivity tensor, 
and $T$ is the absolute temperature. 
The Seebeck tensor $\bm S$ is related to $\bm\Pi$ by the Kelvin-Onsager relation in a magnetic field $\bm{H}$, $\boldsymbol\Pi(\bm{H})=T\boldsymbol S^{\top}(-\bm{H})$, 
which expresses reciprocity between the two conversion processes.
Accordingly, an off-diagonal Seebeck coefficient $S_{yx}$, which produces an electric field along $y$ from
a temperature gradient along $x$, has a reciprocal Peltier coefficient
$\Pi_{xy}$ describing a heat current along $x$ driven by a charge current
along $y$. Such transverse conversion separates the directions of electrical
and thermal transport and provides device geometries distinct from those of
conventional longitudinal thermoelectrics
\cite{Goldsmid2011,Zhou2013,Uchida2022,Adachi2025}.

The best-established transverse thermoelectric phenomena belong to the
Nernst-Ettingshausen family. 
Thermomagnetic cooling based on this mechanism was investigated early in
Bi-Sb alloys \cite{Cuff1963}. 
In magnetic materials,
magnetization and Berry curvature give rise to the anomalous Nernst effect, 
allowing large transverse thermopower even at zero external field
as reported in noncollinear antiferromagnets,
ferromagnetic semimetals, kagome Weyl magnets, and iron-based ferromagnets
\cite{Ikhlas2017,Sakai2018,Guin2019,Sakai2020}.
Its reciprocal process, the
anomalous Ettingshausen effect, has been directly imaged by lock-in
thermography (LIT) \cite{Seki2018,Miura2019}.
However, Nernst/Ettingshausen effect intrinsically requires either an external magnetic field or
magnetic order.

A different route originates from the anisotropy of the thermoelectric tensor and
does not require broken time-reversal symmetry \cite{Adachi2025}. 
An anisotropic diagonal tensor acquires off-diagonal components 
when the measurement axes are rotated relative to the crystallographic axes. 
The resulting off-diagonal Seebeck and Peltier effects have been
investigated in textured YBa$_2$Cu$_3$O$_{7-\delta}$ and in artificially
tilted multilayer systems~\cite{He1996,He1998,Kyarad2006,Reitmaier2010,Uchida2023,Hirai2024}.
These previous realizations, however,
rely either on the crystallographic texturing of an anisotropic material or
on artificially tilted interfaces between dissimilar constituents.
It therefore remains desirable to realize and directly visualize the same
functionality in a homogeneous nonmagnetic single crystal without relying
on crystallographic texturing or engineered internal interfaces.

Axis-dependent conduction polarity (ADCP),
characterized by opposite signs of the Seebeck coefficients along two principal crystallographic axes,
provides an intrinsic realization of this concept \cite{Wang2020,Adachi2025,Song2025,Yang2026,Hu2026}. 
Within this class, $(p\times n)$-type transverse thermoelectricity describes a multiband mechanism in which $p$-type and $n$-type bands dominate transport along different axes \cite{Zhou2013,Grayson2018}, 
whereas goniopolarity describes a single-band mechanism associated with a hyperboloidal open Fermi surface \cite{He2019}.
When the measurement
coordinates $(x,y)$ are rotated by an angle $\theta$ from the crystallographic axes $(i,j)$, 
the transverse Seebeck coefficient in the rotated frame is given as
$S_{yx}=S_{xy}=(S_{ii}-S_{jj})\sin\theta\cos\theta$,
where $S_{ii}$ and $S_{jj}$ are the principal-axis Seebeck coefficients \cite{Scudder2021}.
Thus, the opposite signs of $S_{ii}$ and $S_{jj}$ enhance their difference and hence the
transverse response. 
Note that, since the thermoelectric tensor of the present
nonmagnetic system is symmetric, $S_{xy}=S_{yx}$, 
the reciprocity reduces to
$\bm{\Pi}=T\bm{S}$, 
and the transverse components satisfy
$\Pi_{yx}=TS_{xy}=TS_{yx}$.

Here, we demonstrate both transverse Seebeck and Peltier effects in a semimetal MoSi$_2$,
which exhibits a distinct ADCP property of the Seebeck coefficients of 
$S_{aa}\approx 8$~$\mu$V/K and $S_{cc}\approx -13$~$\mu$V/K at room temperature \cite{Manako2026,Pavlosiuk2026}.
This ADCP originates from a mixed dimensionality of the Fermi surfaces,
coexistence of quasi-one-dimensional and quasi-two-dimensional Fermi surfaces with opposite carrier polarities,
as is proposed in several semimetals \cite{Koster2023,Manako2024,Ohsumi2024,Pavlosiuk2026}.
MoSi$_2$ is therefore classified as a multiband $(p \times n)$-type ADCP material rather than a single-band goniopolar material.
We evaluate the transverse Seebeck response in a MoSi$_2$ single crystal 
by measuring the angular dependence of the 
thermoelectric field under a temperature gradient, and 
find a substantial rotation of the field vector with respect to the temperature gradient vector.
For the Peltier response,
we measured the amplitude and phase images of the MoSi$_2$ single crystal
under an ac electrical current 
by means of harmonic LIT technique,
and find a temperature gradient induced perpendicular to the electrical current.
Since the ADCP in the Seebeck coefficients has been established in MoSi$_2$ \cite{Manako2026,Pavlosiuk2026}, 
the present study addresses a different question: 
it reconstructs the in-plane thermoelectric-field direction using multiple voltage-contact pairs and images the current-odd transverse thermal response on the same crystal.

\begin{figure}[t]
\centering
\includegraphics[width=.5\textwidth]{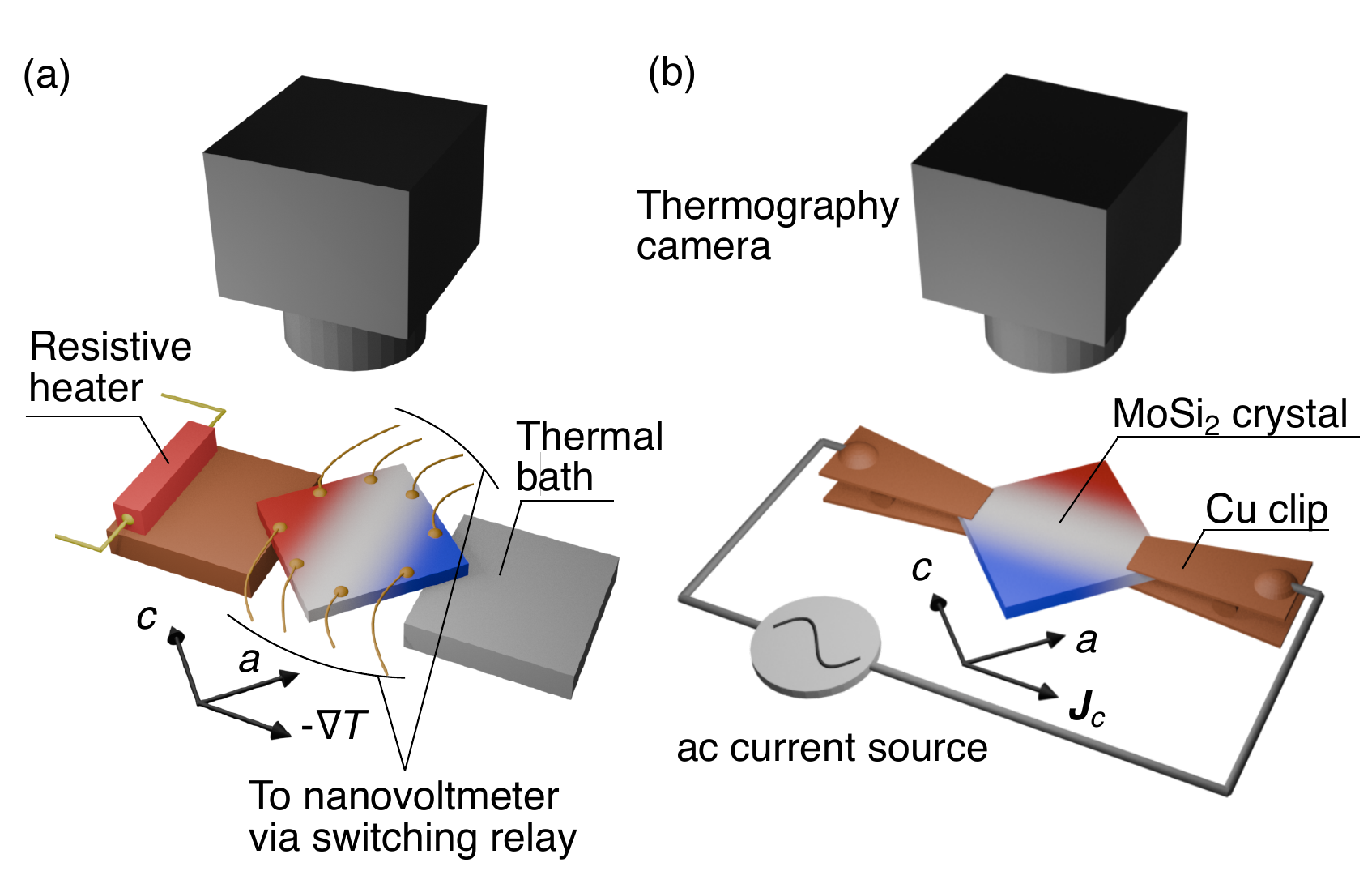}
\caption{
(a) Schematic of the experimental setup for the transverse Seebeck effect.
The crystal is bridged between the resistive heater and thermal bath.
A switching relay is used to measure the thermoelectric voltage between the various terminals of the crystal, which has multiple electrical contacts attached to it.
(b) Schematic of the experimental setup for the transverse Peltier effect.
The crystal is connected to the ac current source through low-contact-resistance copper clips.
The temperature image is then measured by a thermography camera located on the top of the crystal.
}
\label{fig:setup}
\end{figure}

MoSi$_2$ single-crystalline samples were grown from the polycrystalline material 
using the Czochralski method.
The details are described in Ref. \onlinecite{Manako2026}.
The orientation of the single crystals was determined using a Laue camera (Photonic Science, X-ray Laue Back-Scattered Camera).
For the transverse Seebeck measurement,
as schematically shown in Fig.~\ref{fig:setup}(a),
the crystal was mounted between the heater and the thermal bath, 
and 
multiple electrical contacts were attached to its surface.
The temperature gradient was applied in the direction of 
approximately 45-degree rotation from the $a$-axis to the $c$-axis,
and then the thermoelectric voltage was measured 
by sequentially selecting different pairs of voltage contacts with a switching relay.
In the transverse Seebeck measurement, 
a black-body radiation tape is attached on the crystal surface 
to calibrate the infrared emissivity, 
and the applied temperature gradient was measured
using an infrared thermography camera (FSV-1200, Apiste).
The schematic experimental setup 
for the transverse Peltier effect 
is illustrated in Fig.~\ref{fig:setup}(b).
A sinusoidal electrical current $I(t)=I_0\sin(2\pi f t)$ was applied to the crystal 
in the direction of approximately 45-degree rotation from the $a$-axis to the $c$-axis,
and the amplitude and phase images were obtained for fundamental $f$ and the second harmonic $2f$ frequencies.
An insulating black-body coating was applied in this setup.
To examine the experimental results,
a two-dimensional frequency-domain electrothermal finite-element-method (FEM) calculation 
was also performed \cite{Antonova}. 
The complex temperature data was calculated from
$C\frac{\partial T}{\partial t}-\nabla\cdot(\boldsymbol{\kappa}\nabla T) = \bm J_{\rm c}\cdot \bm E
-\nabla\cdot(\boldsymbol\Pi\bm J_{\rm c})$,
where $C$ is the heat capacity \cite{Bose1992}.
Since the sample is heated up under the current (the mean temperature of 320 K for $I=3$~A),
we used the  materials parameters extrapolated to $T=320$~K \cite{Manako2026}.
An isothermal boundary condition is applied 
to both ends of the copper clip.
For other boundaries, we apply the Robin boundary condition \cite{Paloka2024}.
The full calculation used the rotated tensor $\boldsymbol\Pi = T\bm S$ of MoSi$_2$ and the
Peltier coefficient of copper.
On the other hand, the Cu/MoSi$_2$ interface-only calculation with $\Pi_{xy} = \Pi_{yx} = 0$ in MoSi$_2$ 
was also performed for reference (Fig. S6 in Supplementary Material).

\begin{figure}[t]
\centering
\includegraphics[width=.48\textwidth]{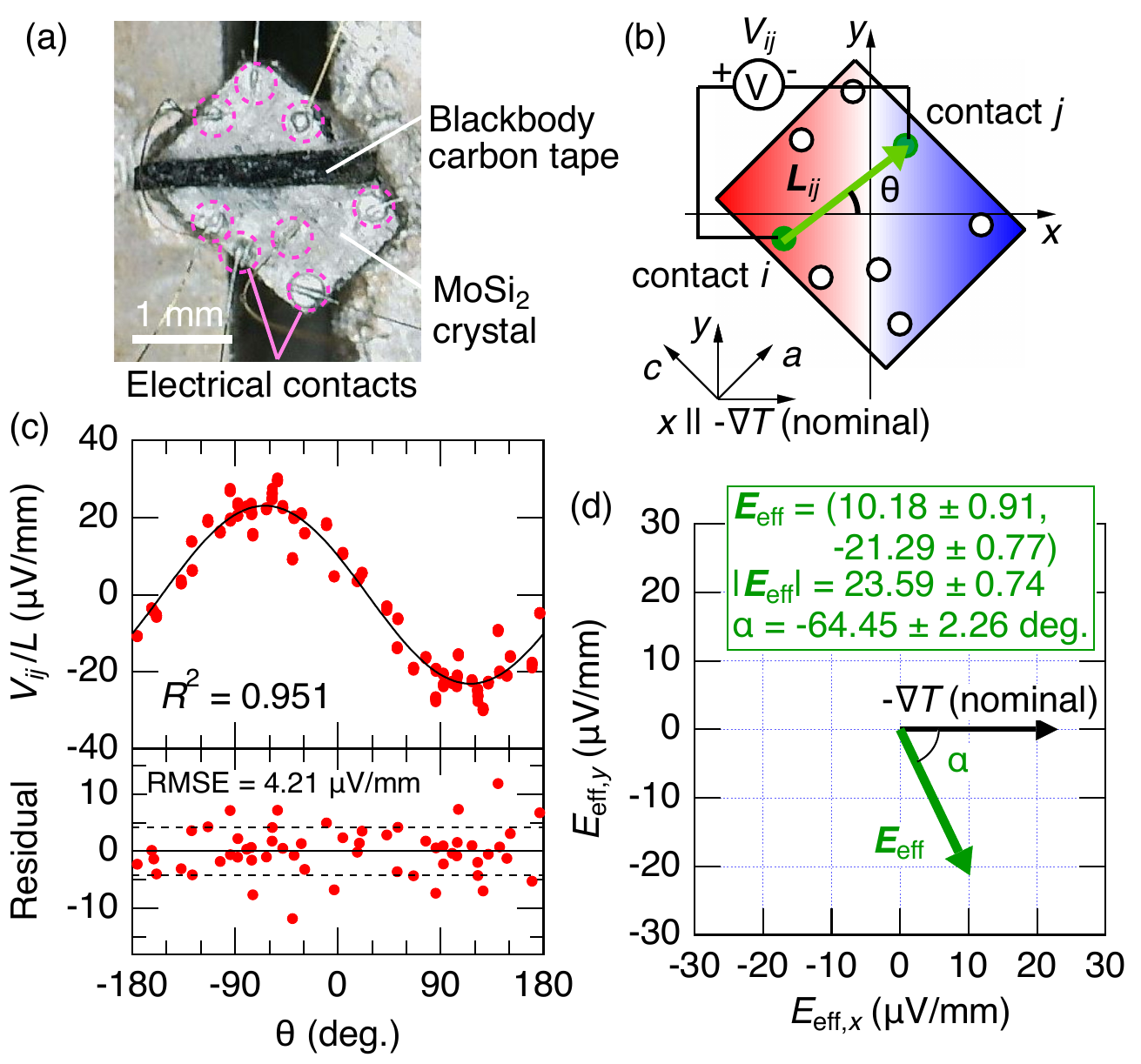}
\caption{
(a) Photograph for the experiment of the transverse Seebeck effect. 
The MoSi$_2$ crystal is bridged between the heater stage and thermal bath.
Eight electrical terminals are attached to the crystal as indicated by pink dashed circles.
On the crystal surface, a black-body radiation tape is attached for calibration.
(b) Schematic experimental configuration. See text for details.
(c) Angular dependence of the thermoelectric field in MoSi$_2$ single crystal measured at room temperature.
The vertical axis represents the electric field approximately given as $V_{ij}/L$, where 
$V_{ij}$ and $L$ are the potential difference and the length between measured contact points, respectively.
The horizontal axis represents the angle measured from the $x$ direction (nominal direction of $-\nabla T$) as shown in (b).
Solid curve is a fitting result shown in the main text.
The goodness-of-fit value is $R^2=0.951$ and the root-mean-square error (RMSE) is 4.21~$\mu$V/mm.
Bottom panel shows the residuals.
(d) Effective thermoelectric-field vector $\bm E_{\rm eff}$ in a Cartesian coordinate.
The nominal direction of $-\nabla T$ is also shown.}
\label{fig:angle}
\end{figure}


We first present the transverse Seebeck effect in a MoSi$_2$ single crystal.
As shown in Figs.~\ref{fig:angle}(a) and \ref{fig:angle}(b),
we applied $-\nabla T$ nominally along the $x$ axis, 
defined as a direction of 45-degrees rotation from the $a$-axis to the $c$-axis,
and measured the potential difference between various terminals
to reconstruct the direction of the thermoelectric field.
The voltage was measured for all ordered pairs of the eight electrical contacts as shown in Figs.~\ref{fig:angle}(a) and \ref{fig:angle}(b).
For each geometrical contact pair $\{i,j\}$, 
two relay configurations were measured separately:
the voltage $V_{ij}$ was measured with
contact $i$ connected to the positive voltmeter input and 
contact $j$ connected to the negative input, 
and 
the voltage $V_{ji}$ was measured with
the polarity-reversed configuration. 
The displacement vector $\bm L_{ij} = L(\cos\theta,\sin\theta)$ 
[shown in Fig.~\ref{fig:angle}(b)] is defined as pointing 
from the positive-input contact $i$ to the negative-input contact $j$. 
The angle $\theta$ 
is therefore
measured from the $x$ direction (nominal direction of $-\nabla T$)
and $L~(>0)$ is the distance between the contact pair $\{i,j\}$.
Thus, reversing the connections gives 
$\bm{L}_{ij}=-\bm{L}_{ji}$ and changes its angle by $180^\circ$, 
while the measured voltage ideally changes sign as $V_{ji}=-V_{ij}$.
Accordingly, the dataset contains $28\times2=56$ ordered measurements.
Detailed datasets are provided in Figs.~S2 and S3 (Supplementary Material).
Measuring both polarities allows us to verify the expected voltage 
reversal and to assess possible polarity-independent instrumental offsets.

Figure~\ref{fig:angle}(c) shows the angular dependence of 
the effective thermoelectric field $V_{ij}/L$.
In the present configuration, 
the observed voltage $V_{ij}$ is approximately expressed as $V_{ij} = \bm E_{\rm eff}\cdot \bm L_{ij}$
under assumption of homogeneous temperature gradient, 
where $\bm E_{\rm eff}$ is the effective electric field, meaning that 
the angle that yields the maximum $V_{ij}/L$ is the direction of the effective electric field.
Thus, the present multi-terminal method allows us to experimentally determine the direction of the field vector.
The data in Fig.~\ref{fig:angle}(c) is fitted with 
\begin{align}
\dfrac{V_{ij}}{L} = 
E_{{\rm eff},x}\cos\theta + 
E_{{\rm eff},y}\sin\theta+
C,
\label{eq:anglerev}
\end{align}
where $\bm E_{\rm eff} = (E_{{\rm eff},x},E_{{\rm eff},y})$ and $C$ is an offset.
The fitting gives $(E_{{\rm eff},x},E_{{\rm eff},y})=(10.18\pm0.91,-21.29\pm0.77)$ and 
$C = -0.03 \pm 0.58$ in $\mu$V/mm, and 
the field vector $\bm E_{\rm eff}$
is directed toward $\alpha = -64.45\pm 2.26$ degrees in this setup,
as is presented in Fig.~\ref{fig:angle}(d).
This result indicates that 
the field vector is significantly rotated from the temperature-gradient direction, 
demonstrating that the 
transverse Seebeck conversion occurs predominantly in MoSi$_2$.

The effective thermoelectric field 
when rotated by an angle $\beta$ from the $a$-axis to the $c$-axis
is given as
\begin{align}
E_{{\rm eff},x} &= (S_{aa}\cos^2\beta + S_{cc}\sin^2\beta -S_{\rm Au})\nabla_xT,
\\
E_{{\rm eff},y} &= [(S_{aa}-S_{cc})\cos\beta\sin\beta]\nabla_xT,
\label{eq:anglerev3}
\end{align}
where 
$S_{aa}$ and $S_{cc}$ are the Seebeck coefficients along the $a$- and $c$-axis directions, respectively,
$S_{\rm Au}$ is the Seebeck coefficient of the gold leads
(approximately $2$~$\mu$V/K at room temperature),
and $\nabla_xT$ ($\approx$ $-2.5$~K/mm) is the temperature gradient along the $x$ axis, which was evaluated by using the thermography camera as shown in Fig.~S1 (Supplementary Material).
The rotation angle in the present experiment is $\beta = \pi/4 +\delta$,
where $\delta \approx 2$~degrees is a small misorientation angle from the 45 degrees.
In this case, the field vector is approximately given as
\begin{align}
E_{{\rm eff},x} &= \dfrac{1}{2}[S_{aa}(1-2\delta) + S_{cc}(1+2\delta) -2S_{\rm Au}]\nabla_xT,
\\
E_{{\rm eff},y} &= \dfrac{1}{2}(S_{aa}-S_{cc})(1-2\delta^2)\nabla_xT.
\label{eq:anglerev4}
\end{align}
The obtained tensor components are 
$S_{aa}\approx 7$~$\mu$V/K and $S_{cc}\approx -11$~$\mu$V/K, 
which are close to the values in the earlier report \cite{Manako2026,Pavlosiuk2026}.

\begin{figure*}[t]
\centering
\includegraphics[width=.99\textwidth]{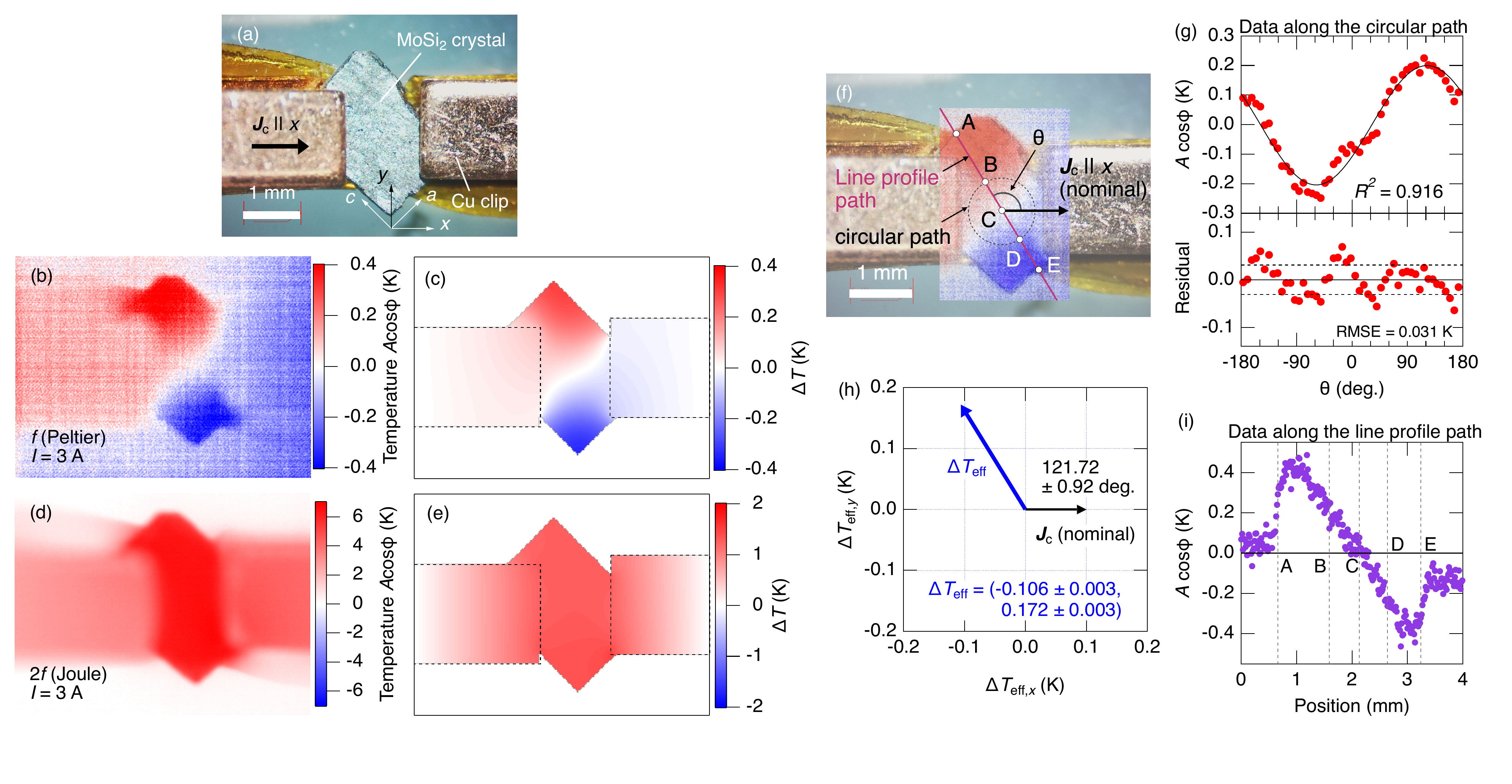}
\caption{(a) Photograph for the experiment of the transverse Peltier effect.
(b) Temperature variation $A\cos\phi$ for the fundamental frequency ($f$-component).
(c) Finite-element-method (FEM) calculation result for the fundamental frequency.
(d) $A\cos\phi$ for the second-harmonic frequency ($2f$-component).
(e) FEM result for the second-harmonic frequency.
(f) Photograph with $f$-component image overlaid.
(g) Angular dependence of local $A\cos\phi$ sampled along the
dashed circular path shown in (f).
Solid curve is a fitting result shown in the main text.
The goodness-of-fit value is $R^2=0.916$ and the RMSE is 0.031 K.
Bottom panel shows the residuals.
(h) 
Effective temperature-variation vector $\Delta T_{\rm eff}$ in
a Cartesian coordinate. The nominal direction of $\bm J_{\rm c}$ is also shown.
(i)
The line profile of local $A\cos\phi$ sampled along 
the direction of $\Delta T_{\rm eff}$ as shown by
the purple solid line in (f).
Points A and E represent the crystal edges, while 
point C is the center of the circular path.
Points B and D represent the points where the line and the circle intersect.
}
\label{fig:P3A}
\end{figure*}

Having established the transverse Seebeck effect in MoSi$_2$, 
let us now  turn to the reciprocal transverse Peltier effect.
The measured crystal is the same one used in the transverse Seebeck measurement
and the photograph of the experimental setup is shown in Fig.~\ref{fig:P3A}(a).
Here we applied sinusoidal electrical current of $I(t) = I_0\sin(2\pi f t)$ with $I_0=3$~A and $f=0.02$~Hz to the crystal nominally
along the $x$ direction, 
defined as a direction of 45-degrees rotation from the $a$-axis to the $c$-axis,
through the copper clips,
and then measured the amplitude $A(\bm r)$ and the phase $\phi(\bm r)$ relative to the applied current
in the 
temperature image $T(\bm r,t) = A(\bm r)\sin[2\pi f' t+\phi(\bm r)]$, where $f'$ is a lock-in frequency.

In Fig.~\ref{fig:P3A}(b), we show the fundamental-frequency ($f'=f$) LIT temperature image 
at the positive maximum of the applied current as
$A(\bm r)\sin[\pi/2+\phi(\bm r)] = A(\bm r)\cos\phi(\bm r)$.
The amplitude and phase images are provided in Fig.~S4 (Supplementary Material).
The fundamental-frequency images represent the Peltier response 
because 
the temperature modulation generated by the Peltier effect $\Delta T_{\rm P}$ is given as 
\begin{align}
\Delta T_{\rm P} \propto\Pi I = \Pi I_0\sin(2\pi ft),
\label{eq:pel}
\end{align}
where $\Pi$ is 
the effective Peltier coefficient for the present geometry.
A pronounced temperature contrast develops along the vertical $y$ direction,
whereas the electrical current is applied nominally along the horizontal $x$ direction.
Opposite temperature modulations appear near the upper and lower edges,
indicating the transverse Peltier effect in MoSi$_2$.
This result is qualitatively reproduced by the FEM calculation shown in Fig.~\ref{fig:P3A}(c),
which includes both 
the transverse Peltier effect in MoSi$_2$ and
the ordinary Peltier effect at the Cu/MoSi$_2$ interfaces.
Note that the ordinary Peltier effect at the Cu/MoSi$_2$ interfaces is relatively small compared with the transverse effect (Fig. S6 in Supplementary Material).
Also,
we observe a non-uniform distribution of the electrical current density 
in the FEM calculation (Fig.~S5 in Supplementary Material).
On the other hand, the direction of the electrical current density in
the central region of the crystal is almost parallel to the $x$ direction and the magnitude of the current
density is small at the edge regions of the crystal, indicating that the impact of
current bending is expected to be relatively minor.

In contrast, 
the second-harmonic ($f'=2f$) image [Fig.~\ref{fig:P3A}(d)] represents the Joule-heating response,
since
the temperature change due to the
Joule heating $\Delta T_{\rm J}$ is given as 
\begin{align}
\Delta T_{\rm J}\propto R I^2 = \dfrac{R I_0^2}{2}[1-\cos(4\pi ft)],
\label{eq:joule}
\end{align}
where $R$ is the resistance.
Although 
the FEM result [Fig.~\ref{fig:P3A}(e)] qualitatively reproduces the experimental data,
the temperature increase is small, which may originate from a 
contact resistance between the copper clip and the crystal, which has not been accounted for here.

To quantify the orientation of the fundamental-frequency temperature pattern, 
we sampled the temperature modulation $A\cos\phi$ 
along the dashed circular path with a radius of 0.54 mm as shown in the overlaid image [Fig.~\ref{fig:P3A}(f)].
Figure~\ref{fig:P3A}(g) shows the angular profile of the sampled temperature change along the circle,
where the angle $\theta$ is measured from the $x$ direction (nominal direction of $\bm J_{\rm c}$),
and 
each plotted value is the mean local $A\cos\phi$ value within one angular bin.
The temperature modulation is fitted with  
\begin{align}
A\cos\phi = 
\Delta T_{{\rm eff},x}\cos\theta + 
\Delta T_{{\rm eff},y}\sin\theta+
C,
\label{eq:Tanglerev}
\end{align}
where $\Delta T_{\rm eff} = (\Delta T_{{\rm eff},x},\Delta T_{{\rm eff},y})$ is 
an effective temperature-variation vector to show the temperature gradient
and $C$ is an offset.
The fitting gives $(\Delta T_{{\rm eff},x},\Delta T_{{\rm eff},y})=(-0.106\pm0.003,0.172\pm0.003)$ and 
$C = 0.002 \pm 0.002$ in K, and 
the vector $\Delta T_{\rm eff}$
is directed toward $121.72\pm 0.92$ degrees in this setup,
as is presented in Fig.~\ref{fig:P3A}(h).
Figure~\ref{fig:P3A}(i) represents the line profile along the direction of the effective temperature-variation vector, showing almost linear temperature variation within the crystal.
The significant rotation of temperature gradient vector $\nabla T$ 
relative to the direction of the current density $\bm J_{\rm c}$ 
demonstrates the transverse Peltier effect in MoSi$_2$,
which is also supported by the FEM calculation result shown in Fig.~\ref{fig:P3A}(c).

\begin{figure}[t!]
\centering
\includegraphics[width=.38\textwidth]{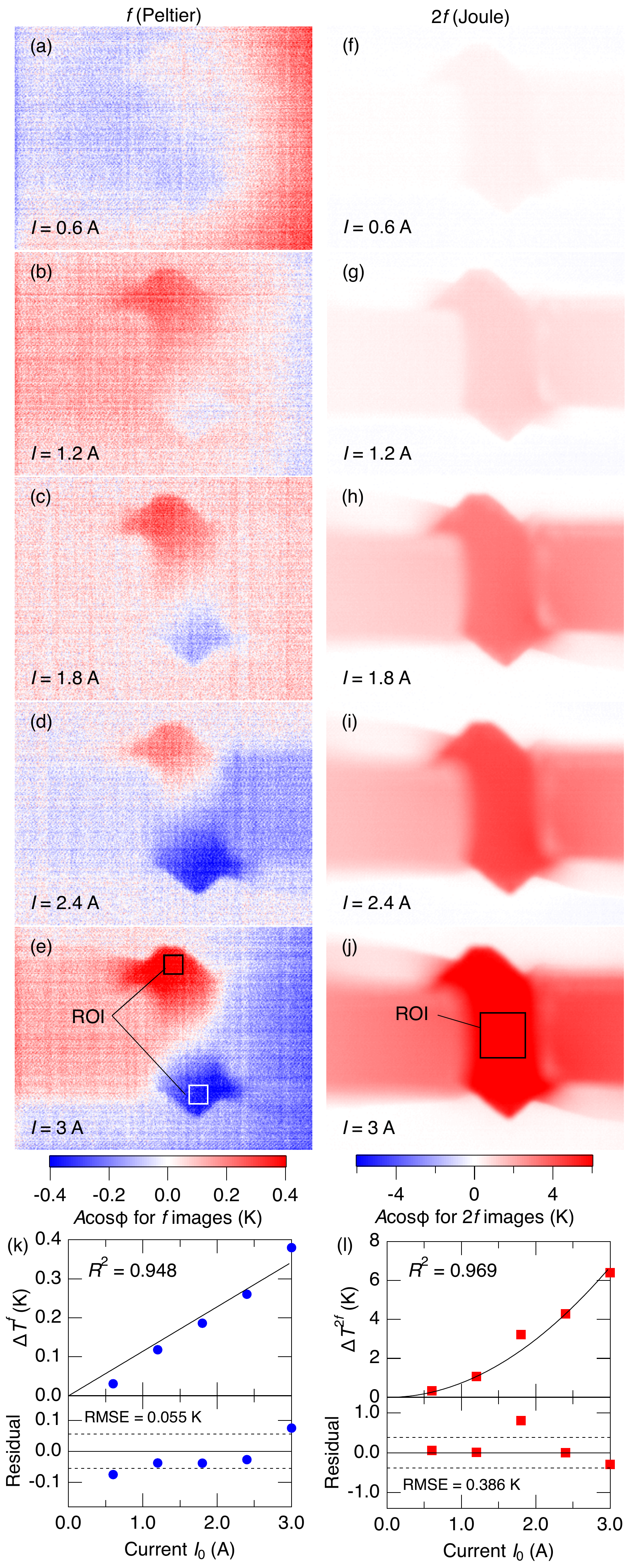}
\caption{
(a-e) The fundamental-frequency ($f$-component) LIT images 
of the temperature variation $A\cos\phi$ measured for various currents.
(f-j) The second-harmonic ($2f$-component) LIT images 
of $A\cos\phi$.
(k,l) Applied current dependence of the local temperature variation for (k) $f$ (Peltier) and (l) $2f$ (Joule) images.
The local temperature variations are determined as an area average
indicated by the region of interest (ROI) in the insets of (e) and (j).
The fitting curves show the linear and quadratic dependencies on current.
The goodness-of-fit value $R^2$, the RMSE, and the residuals are also shown.}
\label{fig:mapcurrent}
\end{figure}

The current dependence of the temperature-variation images
for the $f$ and $2f$ frequencies are shown 
in Figs.~\ref{fig:mapcurrent}(a-e) and \ref{fig:mapcurrent}(f-j), respectively.
In both frequencies, the magnitude increases with increasing current.
Note that small temperature variations in areas other than the crystal and the copper clips may be considered noise.
Figures~\ref{fig:mapcurrent}(k) and \ref{fig:mapcurrent}(l) show the local temperature variations $\Delta T$ in the temperature-variation images
as a function of applied-current amplitude. 
The local temperature variation is determined as an area average in the box shown in the insets of Figs.~\ref{fig:mapcurrent}(e) and \ref{fig:mapcurrent}(j).
The local temperature variation in the $f$ image is determined as $\Delta T^{f} = (\Delta T_{\rm h} - \Delta T_{\rm c})/2$,
where $\Delta T_{\rm h}$ and $\Delta T_{\rm c}$ represent the temperature changes in the warmed and cooled regions, respectively.
The $f$ and $2f$ data are attributable to the Peltier and Joule effects, 
and as expected, exhibit roughly linear and quadratic dependencies on current, respectively.

In summary, 
we demonstrate the transverse Seebeck and Peltier effects
in a single crystal of semimetallic MoSi$_2$,
in which 
the axis-dependent conduction polarity of the Seebeck coefficients is realized.
We first examine the transverse Seebeck effect by utilizing the multi-terminal method 
to evaluate the angular dependence of the thermoelectric field, 
and found a large angle between the temperature gradient and the 
electric field vectors.
We then measure the temperature images under the electrical current by using the lock-in thermography.
The temperature-variation image 
reveals a temperature pattern consistent with a transverse Peltier effect.
Thus, the present results show a direct transverse interconversion between heat and electrical currents in a homogeneous single material.

\section*{Supplementary Material}

Supplementary Material includes 
the thermography image, temperature profile,
contact coordinates, and raw voltage data for the transverse Seebeck measurement. 
For the transverse Peltier measurement,
the amplitude and phase images in the LIT experiment 
and the FEM calculation results are included.
 
\begin{acknowledgments}
We thank H. Manako for contributions during the early stage of this work.
This work was partly supported by JSPS KAKENHI Grant No. 24K06945, 
a research grant from Murata Science and Education Foundation (M25AN115),
and Research Foundation for the Electrotechnology of Chubu (REFEC, No. R-07203).
\end{acknowledgments}

\section*{Data Availability Statement}

The data are available from the authors upon reasonable request.

%


\begin{thebibliography}{32}%
\makeatletter
\providecommand \@ifxundefined [1]{%
 \@ifx{#1\undefined}
}%
\providecommand \@ifnum [1]{%
 \ifnum #1\expandafter \@firstoftwo
 \else \expandafter \@secondoftwo
 \fi
}%
\providecommand \@ifx [1]{%
 \ifx #1\expandafter \@firstoftwo
 \else \expandafter \@secondoftwo
 \fi
}%
\providecommand \natexlab [1]{#1}%
\providecommand \enquote  [1]{``#1''}%
\providecommand \bibnamefont  [1]{#1}%
\providecommand \bibfnamefont [1]{#1}%
\providecommand \citenamefont [1]{#1}%
\providecommand \href@noop [0]{\@secondoftwo}%
\providecommand \href [0]{\begingroup \@sanitize@url \@href}%
\providecommand \@href[1]{\@@startlink{#1}\@@href}%
\providecommand \@@href[1]{\endgroup#1\@@endlink}%
\providecommand \@sanitize@url [0]{\catcode `\\12\catcode `\$12\catcode
  `\&12\catcode `\#12\catcode `\^12\catcode `\_12\catcode `\%12\relax}%
\providecommand \@@startlink[1]{}%
\providecommand \@@endlink[0]{}%
\providecommand \url  [0]{\begingroup\@sanitize@url \@url }%
\providecommand \@url [1]{\endgroup\@href {#1}{\urlprefix }}%
\providecommand \urlprefix  [0]{URL }%
\providecommand \Eprint [0]{\href }%
\providecommand \doibase [0]{https://doi.org/}%
\providecommand \selectlanguage [0]{\@gobble}%
\providecommand \bibinfo  [0]{\@secondoftwo}%
\providecommand \bibfield  [0]{\@secondoftwo}%
\providecommand \translation [1]{[#1]}%
\providecommand \BibitemOpen [0]{}%
\providecommand \bibitemStop [0]{}%
\providecommand \bibitemNoStop [0]{.\EOS\space}%
\providecommand \EOS [0]{\spacefactor3000\relax}%
\providecommand \BibitemShut  [1]{\csname bibitem#1\endcsname}%
\let\auto@bib@innerbib\@empty
\bibitem [{\citenamefont {Goldsmid}(2011)}]{Goldsmid2011}%
  \BibitemOpen
  \bibfield  {author} {\bibinfo {author} {\bibfnamefont {H.~J.}\ \bibnamefont
  {Goldsmid}},\ }\bibfield  {title} {\enquote {\bibinfo {title} {{Application
  of the Transverse Thermoelectric Effects}},}\ }\href
  {https://doi.org/10.1007/s11664-010-1357-3} {\bibfield  {journal} {\bibinfo
  {journal} {J. Electron. Mater.}\ }\textbf {\bibinfo {volume} {40}},\ \bibinfo
  {pages} {1254--1259} (\bibinfo {year} {2011})}\BibitemShut {NoStop}%
\bibitem [{\citenamefont {Zhou}\ \emph {et~al.}(2013)\citenamefont {Zhou},
  \citenamefont {Birner}, \citenamefont {Tang}, \citenamefont {Heinselman},\
  and\ \citenamefont {Grayson}}]{Zhou2013}%
  \BibitemOpen
  \bibfield  {author} {\bibinfo {author} {\bibfnamefont {C.}~\bibnamefont
  {Zhou}}, \bibinfo {author} {\bibfnamefont {S.}~\bibnamefont {Birner}},
  \bibinfo {author} {\bibfnamefont {Y.}~\bibnamefont {Tang}}, \bibinfo {author}
  {\bibfnamefont {K.}~\bibnamefont {Heinselman}},\ and\ \bibinfo {author}
  {\bibfnamefont {M.}~\bibnamefont {Grayson}},\ }\bibfield  {title} {\enquote
  {\bibinfo {title} {{Driving Perpendicular Heat Flow: $(p\times n)$-Type
  Transverse Thermoelectrics for Microscale and Cryogenic Peltier Cooling}},}\
  }\href {https://doi.org/10.1103/PhysRevLett.110.227701} {\bibfield  {journal}
  {\bibinfo  {journal} {Phys. Rev. Lett.}\ }\textbf {\bibinfo {volume} {110}},\
  \bibinfo {pages} {227701} (\bibinfo {year} {2013})}\BibitemShut {NoStop}%
\bibitem [{\citenamefont {Uchida}\ and\ \citenamefont
  {Heremans}(2022)}]{Uchida2022}%
  \BibitemOpen
  \bibfield  {author} {\bibinfo {author} {\bibfnamefont {K.}~\bibnamefont
  {Uchida}}\ and\ \bibinfo {author} {\bibfnamefont {J.~P.}\ \bibnamefont
  {Heremans}},\ }\bibfield  {title} {\enquote {\bibinfo {title}
  {{Thermoelectrics: From Longitudinal to Transverse}},}\ }\href
  {https://doi.org/10.1016/j.joule.2022.08.016} {\bibfield  {journal} {\bibinfo
   {journal} {Joule}\ }\textbf {\bibinfo {volume} {6}},\ \bibinfo {pages}
  {2240--2245} (\bibinfo {year} {2022})}\BibitemShut {NoStop}%
\bibitem [{\citenamefont {Adachi}\ \emph {et~al.}(2025)\citenamefont {Adachi},
  \citenamefont {Ando}, \citenamefont {Hirai}, \citenamefont {Modak},
  \citenamefont {Grayson},\ and\ \citenamefont {Uchida}}]{Adachi2025}%
  \BibitemOpen
  \bibfield  {author} {\bibinfo {author} {\bibfnamefont {H.}~\bibnamefont
  {Adachi}}, \bibinfo {author} {\bibfnamefont {F.}~\bibnamefont {Ando}},
  \bibinfo {author} {\bibfnamefont {T.}~\bibnamefont {Hirai}}, \bibinfo
  {author} {\bibfnamefont {R.}~\bibnamefont {Modak}}, \bibinfo {author}
  {\bibfnamefont {M.}~\bibnamefont {Grayson}},\ and\ \bibinfo {author}
  {\bibfnamefont {K.}~\bibnamefont {Uchida}},\ }\bibfield  {title} {\enquote
  {\bibinfo {title} {{Fundamentals and advances in transverse
  thermoelectrics}},}\ }\href {https://doi.org/10.35848/1882-0786/adf700}
  {\bibfield  {journal} {\bibinfo  {journal} {Appl. Phys. Express}\ }\textbf
  {\bibinfo {volume} {18}},\ \bibinfo {pages} {090101} (\bibinfo {year}
  {2025})}\BibitemShut {NoStop}%
\bibitem [{\citenamefont {Cuff}\ \emph {et~al.}(1963)\citenamefont {Cuff},
  \citenamefont {Horst}, \citenamefont {Weaver}, \citenamefont {Hawkins},
  \citenamefont {Kooi},\ and\ \citenamefont {Enslow}}]{Cuff1963}%
  \BibitemOpen
  \bibfield  {author} {\bibinfo {author} {\bibfnamefont {K.~F.}\ \bibnamefont
  {Cuff}}, \bibinfo {author} {\bibfnamefont {R.~B.}\ \bibnamefont {Horst}},
  \bibinfo {author} {\bibfnamefont {J.~L.}\ \bibnamefont {Weaver}}, \bibinfo
  {author} {\bibfnamefont {S.~R.}\ \bibnamefont {Hawkins}}, \bibinfo {author}
  {\bibfnamefont {C.~F.}\ \bibnamefont {Kooi}},\ and\ \bibinfo {author}
  {\bibfnamefont {G.~M.}\ \bibnamefont {Enslow}},\ }\bibfield  {title}
  {\enquote {\bibinfo {title} {{The Thermomagnetic Figure of Merit and
  Ettingshausen Cooling in Bi--Sb Alloys}},}\ }\href
  {https://doi.org/10.1063/1.1753817} {\bibfield  {journal} {\bibinfo
  {journal} {Appl. Phys. Lett.}\ }\textbf {\bibinfo {volume} {2}},\ \bibinfo
  {pages} {145--146} (\bibinfo {year} {1963})}\BibitemShut {NoStop}%
\bibitem [{\citenamefont {Ikhlas}\ \emph {et~al.}(2017)\citenamefont {Ikhlas},
  \citenamefont {Tomita}, \citenamefont {Koretsune}, \citenamefont {Suzuki},
  \citenamefont {Nishio-Hamane}, \citenamefont {Arita}, \citenamefont {Otani},\
  and\ \citenamefont {Nakatsuji}}]{Ikhlas2017}%
  \BibitemOpen
  \bibfield  {author} {\bibinfo {author} {\bibfnamefont {M.}~\bibnamefont
  {Ikhlas}}, \bibinfo {author} {\bibfnamefont {T.}~\bibnamefont {Tomita}},
  \bibinfo {author} {\bibfnamefont {T.}~\bibnamefont {Koretsune}}, \bibinfo
  {author} {\bibfnamefont {M.-T.}\ \bibnamefont {Suzuki}}, \bibinfo {author}
  {\bibfnamefont {D.}~\bibnamefont {Nishio-Hamane}}, \bibinfo {author}
  {\bibfnamefont {R.}~\bibnamefont {Arita}}, \bibinfo {author} {\bibfnamefont
  {Y.}~\bibnamefont {Otani}},\ and\ \bibinfo {author} {\bibfnamefont
  {S.}~\bibnamefont {Nakatsuji}},\ }\bibfield  {title} {\enquote {\bibinfo
  {title} {{Large anomalous Nernst effect at room temperature in a chiral
  antiferromagnet}},}\ }\href {https://doi.org/10.1038/nphys4181} {\bibfield
  {journal} {\bibinfo  {journal} {Nat. Phys.}\ }\textbf {\bibinfo {volume}
  {13}},\ \bibinfo {pages} {1085--1090} (\bibinfo {year} {2017})}\BibitemShut
  {NoStop}%
\bibitem [{\citenamefont {Sakai}\ \emph {et~al.}(2018)\citenamefont {Sakai},
  \citenamefont {Mizuta}, \citenamefont {Nugroho}, \citenamefont {Sihombing},
  \citenamefont {Koretsune}, \citenamefont {Suzuki}, \citenamefont {Takemori},
  \citenamefont {Ishii}, \citenamefont {Nishio-Hamane}, \citenamefont {Arita},
  \citenamefont {Goswami},\ and\ \citenamefont {Nakatsuji}}]{Sakai2018}%
  \BibitemOpen
  \bibfield  {author} {\bibinfo {author} {\bibfnamefont {A.}~\bibnamefont
  {Sakai}}, \bibinfo {author} {\bibfnamefont {Y.~P.}\ \bibnamefont {Mizuta}},
  \bibinfo {author} {\bibfnamefont {A.~A.}\ \bibnamefont {Nugroho}}, \bibinfo
  {author} {\bibfnamefont {R.}~\bibnamefont {Sihombing}}, \bibinfo {author}
  {\bibfnamefont {T.}~\bibnamefont {Koretsune}}, \bibinfo {author}
  {\bibfnamefont {M.-T.}\ \bibnamefont {Suzuki}}, \bibinfo {author}
  {\bibfnamefont {N.}~\bibnamefont {Takemori}}, \bibinfo {author}
  {\bibfnamefont {R.}~\bibnamefont {Ishii}}, \bibinfo {author} {\bibfnamefont
  {D.}~\bibnamefont {Nishio-Hamane}}, \bibinfo {author} {\bibfnamefont
  {R.}~\bibnamefont {Arita}}, \bibinfo {author} {\bibfnamefont
  {P.}~\bibnamefont {Goswami}},\ and\ \bibinfo {author} {\bibfnamefont
  {S.}~\bibnamefont {Nakatsuji}},\ }\bibfield  {title} {\enquote {\bibinfo
  {title} {{Giant anomalous Nernst effect and quantum-critical scaling in a
  ferromagnetic semimetal}},}\ }\href
  {https://doi.org/10.1038/s41567-018-0225-6} {\bibfield  {journal} {\bibinfo
  {journal} {Nat. Phys.}\ }\textbf {\bibinfo {volume} {14}},\ \bibinfo {pages}
  {1119--1124} (\bibinfo {year} {2018})}\BibitemShut {NoStop}%
\bibitem [{\citenamefont {Guin}\ \emph {et~al.}(2019)\citenamefont {Guin},
  \citenamefont {Vir}, \citenamefont {Zhang}, \citenamefont {Kumar},
  \citenamefont {Watzman}, \citenamefont {Fu}, \citenamefont {Liu},
  \citenamefont {Manna}, \citenamefont {Schnelle}, \citenamefont {Gooth},
  \citenamefont {Shekhar}, \citenamefont {Sun},\ and\ \citenamefont
  {Felser}}]{Guin2019}%
  \BibitemOpen
  \bibfield  {author} {\bibinfo {author} {\bibfnamefont {S.~N.}\ \bibnamefont
  {Guin}}, \bibinfo {author} {\bibfnamefont {P.}~\bibnamefont {Vir}}, \bibinfo
  {author} {\bibfnamefont {Y.}~\bibnamefont {Zhang}}, \bibinfo {author}
  {\bibfnamefont {N.}~\bibnamefont {Kumar}}, \bibinfo {author} {\bibfnamefont
  {S.~J.}\ \bibnamefont {Watzman}}, \bibinfo {author} {\bibfnamefont
  {C.}~\bibnamefont {Fu}}, \bibinfo {author} {\bibfnamefont {E.}~\bibnamefont
  {Liu}}, \bibinfo {author} {\bibfnamefont {K.}~\bibnamefont {Manna}}, \bibinfo
  {author} {\bibfnamefont {W.}~\bibnamefont {Schnelle}}, \bibinfo {author}
  {\bibfnamefont {J.}~\bibnamefont {Gooth}}, \bibinfo {author} {\bibfnamefont
  {C.}~\bibnamefont {Shekhar}}, \bibinfo {author} {\bibfnamefont
  {Y.}~\bibnamefont {Sun}},\ and\ \bibinfo {author} {\bibfnamefont
  {C.}~\bibnamefont {Felser}},\ }\bibfield  {title} {\enquote {\bibinfo {title}
  {{Zero-Field Nernst Effect in a Ferromagnetic Kagome-Lattice Weyl Semimetal
  Co$_3$Sn$_2$S$_2$}},}\ }\href {https://doi.org/10.1002/adma.201806622}
  {\bibfield  {journal} {\bibinfo  {journal} {Adv. Mater.}\ }\textbf {\bibinfo
  {volume} {31}},\ \bibinfo {pages} {1806622} (\bibinfo {year}
  {2019})}\BibitemShut {NoStop}%
\bibitem [{\citenamefont {Sakai}\ \emph {et~al.}(2020)\citenamefont {Sakai},
  \citenamefont {Minami}, \citenamefont {Koretsune}, \citenamefont {Chen},
  \citenamefont {Higo}, \citenamefont {Wang}, \citenamefont {Nomoto},
  \citenamefont {Hirayama}, \citenamefont {Miwa}, \citenamefont
  {Nishio-Hamane}, \citenamefont {Ishii}, \citenamefont {Arita},\ and\
  \citenamefont {Nakatsuji}}]{Sakai2020}%
  \BibitemOpen
  \bibfield  {author} {\bibinfo {author} {\bibfnamefont {A.}~\bibnamefont
  {Sakai}}, \bibinfo {author} {\bibfnamefont {S.}~\bibnamefont {Minami}},
  \bibinfo {author} {\bibfnamefont {T.}~\bibnamefont {Koretsune}}, \bibinfo
  {author} {\bibfnamefont {T.}~\bibnamefont {Chen}}, \bibinfo {author}
  {\bibfnamefont {T.}~\bibnamefont {Higo}}, \bibinfo {author} {\bibfnamefont
  {Y.}~\bibnamefont {Wang}}, \bibinfo {author} {\bibfnamefont {T.}~\bibnamefont
  {Nomoto}}, \bibinfo {author} {\bibfnamefont {M.}~\bibnamefont {Hirayama}},
  \bibinfo {author} {\bibfnamefont {S.}~\bibnamefont {Miwa}}, \bibinfo {author}
  {\bibfnamefont {D.}~\bibnamefont {Nishio-Hamane}}, \bibinfo {author}
  {\bibfnamefont {F.}~\bibnamefont {Ishii}}, \bibinfo {author} {\bibfnamefont
  {R.}~\bibnamefont {Arita}},\ and\ \bibinfo {author} {\bibfnamefont
  {S.}~\bibnamefont {Nakatsuji}},\ }\bibfield  {title} {\enquote {\bibinfo
  {title} {{Iron-based binary ferromagnets for transverse thermoelectric
  conversion}},}\ }\href {https://doi.org/10.1038/s41586-020-2230-z} {\bibfield
   {journal} {\bibinfo  {journal} {Nature}\ }\textbf {\bibinfo {volume}
  {581}},\ \bibinfo {pages} {53--57} (\bibinfo {year} {2020})}\BibitemShut
  {NoStop}%
\bibitem [{\citenamefont {Seki}\ \emph {et~al.}(2018)\citenamefont {Seki},
  \citenamefont {Iguchi}, \citenamefont {Takanashi},\ and\ \citenamefont
  {Uchida}}]{Seki2018}%
  \BibitemOpen
  \bibfield  {author} {\bibinfo {author} {\bibfnamefont {T.}~\bibnamefont
  {Seki}}, \bibinfo {author} {\bibfnamefont {R.}~\bibnamefont {Iguchi}},
  \bibinfo {author} {\bibfnamefont {K.}~\bibnamefont {Takanashi}},\ and\
  \bibinfo {author} {\bibfnamefont {K.}~\bibnamefont {Uchida}},\ }\bibfield
  {title} {\enquote {\bibinfo {title} {{Visualization of anomalous
  Ettingshausen effect in a ferromagnetic film: Direct evidence of different
  symmetry from spin Peltier effect}},}\ }\href
  {https://doi.org/10.1063/1.5022759} {\bibfield  {journal} {\bibinfo
  {journal} {Appl. Phys. Lett.}\ }\textbf {\bibinfo {volume} {112}},\ \bibinfo
  {pages} {152403} (\bibinfo {year} {2018})}\BibitemShut {NoStop}%
\bibitem [{\citenamefont {Miura}\ \emph {et~al.}(2019)\citenamefont {Miura},
  \citenamefont {Sepehri-Amin}, \citenamefont {Masuda}, \citenamefont
  {Tsuchiura}, \citenamefont {Miura}, \citenamefont {Iguchi}, \citenamefont
  {Sakuraba}, \citenamefont {Shiomi}, \citenamefont {Hono},\ and\ \citenamefont
  {Uchida}}]{Miura2019}%
  \BibitemOpen
  \bibfield  {author} {\bibinfo {author} {\bibfnamefont {A.}~\bibnamefont
  {Miura}}, \bibinfo {author} {\bibfnamefont {H.}~\bibnamefont {Sepehri-Amin}},
  \bibinfo {author} {\bibfnamefont {K.}~\bibnamefont {Masuda}}, \bibinfo
  {author} {\bibfnamefont {H.}~\bibnamefont {Tsuchiura}}, \bibinfo {author}
  {\bibfnamefont {Y.}~\bibnamefont {Miura}}, \bibinfo {author} {\bibfnamefont
  {R.}~\bibnamefont {Iguchi}}, \bibinfo {author} {\bibfnamefont
  {Y.}~\bibnamefont {Sakuraba}}, \bibinfo {author} {\bibfnamefont
  {J.}~\bibnamefont {Shiomi}}, \bibinfo {author} {\bibfnamefont
  {K.}~\bibnamefont {Hono}},\ and\ \bibinfo {author} {\bibfnamefont
  {K.}~\bibnamefont {Uchida}},\ }\bibfield  {title} {\enquote {\bibinfo {title}
  {{Observation of anomalous Ettingshausen effect and large transverse
  thermoelectric conductivity in permanent magnets}},}\ }\href
  {https://doi.org/10.1063/1.5131001} {\bibfield  {journal} {\bibinfo
  {journal} {Appl. Phys. Lett.}\ }\textbf {\bibinfo {volume} {115}},\ \bibinfo
  {pages} {222403} (\bibinfo {year} {2019})}\BibitemShut {NoStop}%
\bibitem [{\citenamefont {He}\ \emph {et~al.}(1996)\citenamefont {He},
  \citenamefont {Ma}, \citenamefont {Li}, \citenamefont {Luo}, \citenamefont
  {Zhang}, \citenamefont {Meng},\ and\ \citenamefont {Chu}}]{He1996}%
  \BibitemOpen
  \bibfield  {author} {\bibinfo {author} {\bibfnamefont {Z.~H.}\ \bibnamefont
  {He}}, \bibinfo {author} {\bibfnamefont {Z.~G.}\ \bibnamefont {Ma}}, \bibinfo
  {author} {\bibfnamefont {Q.~Y.}\ \bibnamefont {Li}}, \bibinfo {author}
  {\bibfnamefont {Y.~Y.}\ \bibnamefont {Luo}}, \bibinfo {author} {\bibfnamefont
  {J.~X.}\ \bibnamefont {Zhang}}, \bibinfo {author} {\bibfnamefont {R.~L.}\
  \bibnamefont {Meng}},\ and\ \bibinfo {author} {\bibfnamefont {C.~W.}\
  \bibnamefont {Chu}},\ }\bibfield  {title} {\enquote {\bibinfo {title}
  {{Investigation of transverse Peltier effect on top-seeded melt texture
  YBa$_2$Cu$_3$O$_{7-\delta}$}},}\ }\href {https://doi.org/10.1063/1.117215}
  {\bibfield  {journal} {\bibinfo  {journal} {Appl. Phys. Lett.}\ }\textbf
  {\bibinfo {volume} {69}},\ \bibinfo {pages} {3587--3589} (\bibinfo {year}
  {1996})}\BibitemShut {NoStop}%
\bibitem [{\citenamefont {He}\ \emph {et~al.}(1998)\citenamefont {He},
  \citenamefont {Ma}, \citenamefont {Zhong}, \citenamefont {Luo}, \citenamefont
  {Li}, \citenamefont {Wu},\ and\ \citenamefont {Gavalek}}]{He1998}%
  \BibitemOpen
  \bibfield  {author} {\bibinfo {author} {\bibfnamefont {Z.~H.}\ \bibnamefont
  {He}}, \bibinfo {author} {\bibfnamefont {Z.~G.}\ \bibnamefont {Ma}}, \bibinfo
  {author} {\bibfnamefont {F.~Q.}\ \bibnamefont {Zhong}}, \bibinfo {author}
  {\bibfnamefont {Y.~Y.}\ \bibnamefont {Luo}}, \bibinfo {author} {\bibfnamefont
  {Q.~Y.}\ \bibnamefont {Li}}, \bibinfo {author} {\bibfnamefont
  {M.}~\bibnamefont {Wu}},\ and\ \bibinfo {author} {\bibfnamefont
  {W.}~\bibnamefont {Gavalek}},\ }\bibfield  {title} {\enquote {\bibinfo
  {title} {{Investigation of the Off-Diagonal Thermoelectric Effect on Textured
  YBa$_2$Cu$_3$O$_{7-\delta}$}},}\ }\href
  {https://doi.org/10.1023/A:1022663313363} {\bibfield  {journal} {\bibinfo
  {journal} {J. Supercond.}\ }\textbf {\bibinfo {volume} {11}},\ \bibinfo
  {pages} {203--207} (\bibinfo {year} {1998})}\BibitemShut {NoStop}%
\bibitem [{\citenamefont {Kyarad}\ and\ \citenamefont
  {Lengfellner}(2006)}]{Kyarad2006}%
  \BibitemOpen
  \bibfield  {author} {\bibinfo {author} {\bibfnamefont {A.}~\bibnamefont
  {Kyarad}}\ and\ \bibinfo {author} {\bibfnamefont {H.}~\bibnamefont
  {Lengfellner}},\ }\bibfield  {title} {\enquote {\bibinfo {title} {{Transverse
  Peltier effect in tilted Pb--Bi$_2$Te$_3$ multilayer structures}},}\ }\href
  {https://doi.org/10.1063/1.2385213} {\bibfield  {journal} {\bibinfo
  {journal} {Appl. Phys. Lett.}\ }\textbf {\bibinfo {volume} {89}},\ \bibinfo
  {pages} {192103} (\bibinfo {year} {2006})}\BibitemShut {NoStop}%
\bibitem [{\citenamefont {Reitmaier}, \citenamefont {Walther},\ and\
  \citenamefont {Lengfellner}(2010)}]{Reitmaier2010}%
  \BibitemOpen
  \bibfield  {author} {\bibinfo {author} {\bibfnamefont {C.}~\bibnamefont
  {Reitmaier}}, \bibinfo {author} {\bibfnamefont {F.}~\bibnamefont {Walther}},\
  and\ \bibinfo {author} {\bibfnamefont {H.}~\bibnamefont {Lengfellner}},\
  }\bibfield  {title} {\enquote {\bibinfo {title} {Transverse thermoelectric
  devices},}\ }\href {https://doi.org/10.1007/s00339-010-5742-9} {\bibfield
  {journal} {\bibinfo  {journal} {Appl. Phys. A}\ }\textbf {\bibinfo {volume}
  {99}},\ \bibinfo {pages} {717--722} (\bibinfo {year} {2010})}\BibitemShut
  {NoStop}%
\bibitem [{\citenamefont {Uchida}\ \emph {et~al.}(2023)\citenamefont {Uchida},
  \citenamefont {Hirai}, \citenamefont {Ando},\ and\ \citenamefont
  {Sepehri-Amin}}]{Uchida2023}%
  \BibitemOpen
  \bibfield  {author} {\bibinfo {author} {\bibfnamefont {K.}~\bibnamefont
  {Uchida}}, \bibinfo {author} {\bibfnamefont {T.}~\bibnamefont {Hirai}},
  \bibinfo {author} {\bibfnamefont {F.}~\bibnamefont {Ando}},\ and\ \bibinfo
  {author} {\bibfnamefont {H.}~\bibnamefont {Sepehri-Amin}},\ }\bibfield
  {title} {\enquote {\bibinfo {title} {{Hybrid Transverse
  Magneto-Thermoelectric Cooling in Artificially Tilted Multilayers}},}\ }\href
  {https://doi.org/10.1002/aenm.202302375} {\bibfield  {journal} {\bibinfo
  {journal} {Adv. Energy Mater.}\ }\textbf {\bibinfo {volume} {14}},\ \bibinfo
  {pages} {2302375} (\bibinfo {year} {2023})}\BibitemShut {NoStop}%
\bibitem [{\citenamefont {Hirai}\ \emph {et~al.}(2024)\citenamefont {Hirai},
  \citenamefont {Ando}, \citenamefont {Sepehri-Amin},\ and\ \citenamefont
  {Uchida}}]{Hirai2024}%
  \BibitemOpen
  \bibfield  {author} {\bibinfo {author} {\bibfnamefont {T.}~\bibnamefont
  {Hirai}}, \bibinfo {author} {\bibfnamefont {F.}~\bibnamefont {Ando}},
  \bibinfo {author} {\bibfnamefont {H.}~\bibnamefont {Sepehri-Amin}},\ and\
  \bibinfo {author} {\bibfnamefont {K.}~\bibnamefont {Uchida}},\ }\bibfield
  {title} {\enquote {\bibinfo {title} {{Hybridizing anomalous Nernst effect in
  artificially tilted multilayer based on magnetic topological material}},}\
  }\href {https://doi.org/10.1038/s41467-024-53723-2} {\bibfield  {journal}
  {\bibinfo  {journal} {Nat. Commun.}\ }\textbf {\bibinfo {volume} {15}},\
  \bibinfo {pages} {9643} (\bibinfo {year} {2024})}\BibitemShut {NoStop}%
\bibitem [{\citenamefont {Wang}\ \emph {et~al.}(2020)\citenamefont {Wang},
  \citenamefont {Koster}, \citenamefont {Ochs}, \citenamefont {Scudder},
  \citenamefont {Heremans}, \citenamefont {Windl},\ and\ \citenamefont
  {Goldberger}}]{Wang2020}%
  \BibitemOpen
  \bibfield  {author} {\bibinfo {author} {\bibfnamefont {Y.}~\bibnamefont
  {Wang}}, \bibinfo {author} {\bibfnamefont {K.~G.}\ \bibnamefont {Koster}},
  \bibinfo {author} {\bibfnamefont {A.~M.}\ \bibnamefont {Ochs}}, \bibinfo
  {author} {\bibfnamefont {M.~R.}\ \bibnamefont {Scudder}}, \bibinfo {author}
  {\bibfnamefont {J.~P.}\ \bibnamefont {Heremans}}, \bibinfo {author}
  {\bibfnamefont {W.}~\bibnamefont {Windl}},\ and\ \bibinfo {author}
  {\bibfnamefont {J.~E.}\ \bibnamefont {Goldberger}},\ }\bibfield  {title}
  {\enquote {\bibinfo {title} {{The Chemical Design Principles for
  Axis-Dependent Conduction Polarity}},}\ }\href
  {https://doi.org/10.1021/jacs.9b10626} {\bibfield  {journal} {\bibinfo
  {journal} {J. Am. Chem. Soc.}\ }\textbf {\bibinfo {volume} {142}},\ \bibinfo
  {pages} {2812--2822} (\bibinfo {year} {2020})}\BibitemShut {NoStop}%
\bibitem [{\citenamefont {Song}\ and\ \citenamefont
  {Goldberger}(2025)}]{Song2025}%
  \BibitemOpen
  \bibfield  {author} {\bibinfo {author} {\bibfnamefont {T.}~\bibnamefont
  {Song}}\ and\ \bibinfo {author} {\bibfnamefont {J.~E.}\ \bibnamefont
  {Goldberger}},\ }\bibfield  {title} {\enquote {\bibinfo {title} {{A Field
  Guide to Materials with Axis-Dependent Conduction Polarity}},}\ }\href
  {https://doi.org/10.1021/acs.chemmater.5c01707} {\bibfield  {journal}
  {\bibinfo  {journal} {Chem. Mater.}\ }\textbf {\bibinfo {volume} {37}},\
  \bibinfo {pages} {7518--7533} (\bibinfo {year} {2025})}\BibitemShut {NoStop}%
\bibitem [{\citenamefont {Yang}\ \emph {et~al.}(2026)\citenamefont {Yang},
  \citenamefont {He}, \citenamefont {Usui}, \citenamefont {Kamiya},\ and\
  \citenamefont {Katase}}]{Yang2026}%
  \BibitemOpen
  \bibfield  {author} {\bibinfo {author} {\bibfnamefont {Z.}~\bibnamefont
  {Yang}}, \bibinfo {author} {\bibfnamefont {X.}~\bibnamefont {He}}, \bibinfo
  {author} {\bibfnamefont {H.}~\bibnamefont {Usui}}, \bibinfo {author}
  {\bibfnamefont {T.}~\bibnamefont {Kamiya}},\ and\ \bibinfo {author}
  {\bibfnamefont {T.}~\bibnamefont {Katase}},\ }\bibfield  {title} {\enquote
  {\bibinfo {title} {{Axis-Dependent Conduction Polarity: Design Principles and
  High-Throughput Discovery of Transverse Thermoelectrics}},}\ }\href
  {https://doi.org/10.1021/jacs.5c22733} {\bibfield  {journal} {\bibinfo
  {journal} {J. Am. Chem. Soc.}\ }\textbf {\bibinfo {volume} {148}},\ \bibinfo
  {pages} {6704--6715} (\bibinfo {year} {2026})}\BibitemShut {NoStop}%
\bibitem [{\citenamefont {Hu}\ \emph {et~al.}(2026)\citenamefont {Hu},
  \citenamefont {Ju}, \citenamefont {Cheng}, \citenamefont {Feng},
  \citenamefont {Sun}, \citenamefont {Lou}, \citenamefont {Schnelle},
  \citenamefont {Koban}, \citenamefont {Wang}, \citenamefont {Fedorov},
  \citenamefont {Suvorov}, \citenamefont {Jana}, \citenamefont {Fujii},
  \citenamefont {Vobornik}, \citenamefont {Vyalikh}, \citenamefont
  {B{\"u}chner}, \citenamefont {He}, \citenamefont {Pan},\ and\ \citenamefont
  {Felser}}]{Hu2026}%
  \BibitemOpen
  \bibfield  {author} {\bibinfo {author} {\bibfnamefont {H.}~\bibnamefont
  {Hu}}, \bibinfo {author} {\bibfnamefont {Y.}~\bibnamefont {Ju}}, \bibinfo
  {author} {\bibfnamefont {E.}~\bibnamefont {Cheng}}, \bibinfo {author}
  {\bibfnamefont {X.}~\bibnamefont {Feng}}, \bibinfo {author} {\bibfnamefont
  {F.}~\bibnamefont {Sun}}, \bibinfo {author} {\bibfnamefont {R.}~\bibnamefont
  {Lou}}, \bibinfo {author} {\bibfnamefont {W.}~\bibnamefont {Schnelle}},
  \bibinfo {author} {\bibfnamefont {R.}~\bibnamefont {Koban}}, \bibinfo
  {author} {\bibfnamefont {H.}~\bibnamefont {Wang}}, \bibinfo {author}
  {\bibfnamefont {A.}~\bibnamefont {Fedorov}}, \bibinfo {author} {\bibfnamefont
  {O.}~\bibnamefont {Suvorov}}, \bibinfo {author} {\bibfnamefont
  {A.}~\bibnamefont {Jana}}, \bibinfo {author} {\bibfnamefont {J.}~\bibnamefont
  {Fujii}}, \bibinfo {author} {\bibfnamefont {I.}~\bibnamefont {Vobornik}},
  \bibinfo {author} {\bibfnamefont {D.~V.}\ \bibnamefont {Vyalikh}}, \bibinfo
  {author} {\bibfnamefont {B.}~\bibnamefont {B{\"u}chner}}, \bibinfo {author}
  {\bibfnamefont {B.}~\bibnamefont {He}}, \bibinfo {author} {\bibfnamefont
  {X.}~\bibnamefont {Pan}},\ and\ \bibinfo {author} {\bibfnamefont
  {C.}~\bibnamefont {Felser}},\ }\bibfield  {title} {\enquote {\bibinfo {title}
  {{Fermiology-driven goniopolar transverse thermoelectricity in kagome
  metals}},}\ }\href {https://doi.org/10.1038/s41563-026-02678-4} {\bibfield
  {journal} {\bibinfo  {journal} {Nat. Mater.}\ } (\bibinfo {year} {2026}),\
  10.1038/s41563-026-02678-4}\BibitemShut {NoStop}%
\bibitem [{\citenamefont {Grayson}\ \emph {et~al.}(2018)\citenamefont
  {Grayson}, \citenamefont {Shao}, \citenamefont {Cui}, \citenamefont {Tang},
  \citenamefont {Yan},\ and\ \citenamefont {Zhou}}]{Grayson2018}%
  \BibitemOpen
  \bibfield  {author} {\bibinfo {author} {\bibfnamefont {M.}~\bibnamefont
  {Grayson}}, \bibinfo {author} {\bibfnamefont {Q.}~\bibnamefont {Shao}},
  \bibinfo {author} {\bibfnamefont {B.}~\bibnamefont {Cui}}, \bibinfo {author}
  {\bibfnamefont {Y.}~\bibnamefont {Tang}}, \bibinfo {author} {\bibfnamefont
  {X.}~\bibnamefont {Yan}},\ and\ \bibinfo {author} {\bibfnamefont
  {C.}~\bibnamefont {Zhou}},\ }\bibfield  {title} {\enquote {\bibinfo {title}
  {{Introduction to $(p\times n)$-Type Transverse Thermoelectrics}},}\ }in\
  \href {https://doi.org/10.5772/intechopen.78718} {\emph {\bibinfo {booktitle}
  {Bringing Thermoelectricity into Reality}}}\ (\bibinfo  {publisher}
  {IntechOpen},\ \bibinfo {year} {2018})\BibitemShut {NoStop}%
\bibitem [{\citenamefont {He}\ \emph {et~al.}(2019)\citenamefont {He},
  \citenamefont {Wang}, \citenamefont {Arguilla}, \citenamefont {Cultrara},
  \citenamefont {Scudder}, \citenamefont {Goldberger}, \citenamefont {Windl},\
  and\ \citenamefont {Heremans}}]{He2019}%
  \BibitemOpen
  \bibfield  {author} {\bibinfo {author} {\bibfnamefont {B.}~\bibnamefont
  {He}}, \bibinfo {author} {\bibfnamefont {Y.}~\bibnamefont {Wang}}, \bibinfo
  {author} {\bibfnamefont {M.~Q.}\ \bibnamefont {Arguilla}}, \bibinfo {author}
  {\bibfnamefont {N.~D.}\ \bibnamefont {Cultrara}}, \bibinfo {author}
  {\bibfnamefont {M.~R.}\ \bibnamefont {Scudder}}, \bibinfo {author}
  {\bibfnamefont {J.~E.}\ \bibnamefont {Goldberger}}, \bibinfo {author}
  {\bibfnamefont {W.}~\bibnamefont {Windl}},\ and\ \bibinfo {author}
  {\bibfnamefont {J.~P.}\ \bibnamefont {Heremans}},\ }\bibfield  {title}
  {\enquote {\bibinfo {title} {{The Fermi surface geometrical origin of
  axis-dependent conduction polarity in layered materials}},}\ }\href
  {https://doi.org/10.1038/s41563-019-0309-4} {\bibfield  {journal} {\bibinfo
  {journal} {Nat. Mater.}\ }\textbf {\bibinfo {volume} {18}},\ \bibinfo {pages}
  {568--572} (\bibinfo {year} {2019})}\BibitemShut {NoStop}%
\bibitem [{\citenamefont {Scudder}\ \emph {et~al.}(2021)\citenamefont
  {Scudder}, \citenamefont {He}, \citenamefont {Wang}, \citenamefont {Rai},
  \citenamefont {Cahill}, \citenamefont {Windl}, \citenamefont {Heremans},\
  and\ \citenamefont {Goldberger}}]{Scudder2021}%
  \BibitemOpen
  \bibfield  {author} {\bibinfo {author} {\bibfnamefont {M.~R.}\ \bibnamefont
  {Scudder}}, \bibinfo {author} {\bibfnamefont {B.}~\bibnamefont {He}},
  \bibinfo {author} {\bibfnamefont {Y.}~\bibnamefont {Wang}}, \bibinfo {author}
  {\bibfnamefont {A.}~\bibnamefont {Rai}}, \bibinfo {author} {\bibfnamefont
  {D.~G.}\ \bibnamefont {Cahill}}, \bibinfo {author} {\bibfnamefont
  {W.}~\bibnamefont {Windl}}, \bibinfo {author} {\bibfnamefont {J.~P.}\
  \bibnamefont {Heremans}},\ and\ \bibinfo {author} {\bibfnamefont {J.~E.}\
  \bibnamefont {Goldberger}},\ }\bibfield  {title} {\enquote {\bibinfo {title}
  {{Highly efficient transverse thermoelectric devices with Re$_4$Si$_7$
  crystals}},}\ }\href {https://doi.org/10.1039/d1ee00923k} {\bibfield
  {journal} {\bibinfo  {journal} {Energy Environ. Sci.}\ }\textbf {\bibinfo
  {volume} {14}},\ \bibinfo {pages} {4009--4017} (\bibinfo {year}
  {2021})}\BibitemShut {NoStop}%
\bibitem [{\citenamefont {Manako}\ \emph {et~al.}(2026)\citenamefont {Manako},
  \citenamefont {Ohsumi}, \citenamefont {Yoshida}, \citenamefont {Okazaki},\
  and\ \citenamefont {Sato}}]{Manako2026}%
  \BibitemOpen
  \bibfield  {author} {\bibinfo {author} {\bibfnamefont {H.}~\bibnamefont
  {Manako}}, \bibinfo {author} {\bibfnamefont {S.}~\bibnamefont {Ohsumi}},
  \bibinfo {author} {\bibfnamefont {S.}~\bibnamefont {Yoshida}}, \bibinfo
  {author} {\bibfnamefont {R.}~\bibnamefont {Okazaki}},\ and\ \bibinfo {author}
  {\bibfnamefont {Y.~J.}\ \bibnamefont {Sato}},\ }\bibfield  {title} {\enquote
  {\bibinfo {title} {{Axis-dependent conduction polarity and transverse
  thermoelectric conversion in the mixed-dimensional semimetal MoSi$_2$}},}\
  }\href {https://doi.org/10.1038/s43246-025-01050-4} {\bibfield  {journal}
  {\bibinfo  {journal} {Commun. Mater.}\ }\textbf {\bibinfo {volume} {7}},\
  \bibinfo {pages} {41} (\bibinfo {year} {2026})}\BibitemShut {NoStop}%
\bibitem [{\citenamefont {Pavlosiuk}\ and\ \citenamefont
  {Winiarski}(2026)}]{Pavlosiuk2026}%
  \BibitemOpen
  \bibfield  {author} {\bibinfo {author} {\bibfnamefont {O.}~\bibnamefont
  {Pavlosiuk}}\ and\ \bibinfo {author} {\bibfnamefont {M.~J.}\ \bibnamefont
  {Winiarski}},\ }\bibfield  {title} {\enquote {\bibinfo {title} {{Coexistence
  of Large Nernst Effect and Axis-Dependent Conduction Polarity in Semimetal
  MoSi$_2$}},}\ }\href {https://doi.org/10.1002/aelm.70563} {\bibfield
  {journal} {\bibinfo  {journal} {Adv. Electron. Mater.}\ } (\bibinfo {year}
  {2026}),\ 10.1002/aelm.70563}\BibitemShut {NoStop}%
\bibitem [{\citenamefont {Koster}\ \emph {et~al.}(2023)\citenamefont {Koster},
  \citenamefont {Deng}, \citenamefont {Moore}, \citenamefont {Heremans},
  \citenamefont {Windl},\ and\ \citenamefont {Goldberger}}]{Koster2023}%
  \BibitemOpen
  \bibfield  {author} {\bibinfo {author} {\bibfnamefont {K.~G.}\ \bibnamefont
  {Koster}}, \bibinfo {author} {\bibfnamefont {Z.}~\bibnamefont {Deng}},
  \bibinfo {author} {\bibfnamefont {C.~E.}\ \bibnamefont {Moore}}, \bibinfo
  {author} {\bibfnamefont {J.~P.}\ \bibnamefont {Heremans}}, \bibinfo {author}
  {\bibfnamefont {W.}~\bibnamefont {Windl}},\ and\ \bibinfo {author}
  {\bibfnamefont {J.~E.}\ \bibnamefont {Goldberger}},\ }\bibfield  {title}
  {\enquote {\bibinfo {title} {{Axis-Dependent Conduction Polarity in WSi$_2$
  Single Crystals}},}\ }\href {https://doi.org/10.1021/acs.chemmater.3c00183}
  {\bibfield  {journal} {\bibinfo  {journal} {Chem. Mater.}\ }\textbf {\bibinfo
  {volume} {35}},\ \bibinfo {pages} {4228--4234} (\bibinfo {year}
  {2023})}\BibitemShut {NoStop}%
\bibitem [{\citenamefont {Manako}\ \emph {et~al.}(2024)\citenamefont {Manako},
  \citenamefont {Ohsumi}, \citenamefont {Sato}, \citenamefont {Okazaki},\ and\
  \citenamefont {Aoki}}]{Manako2024}%
  \BibitemOpen
  \bibfield  {author} {\bibinfo {author} {\bibfnamefont {H.}~\bibnamefont
  {Manako}}, \bibinfo {author} {\bibfnamefont {S.}~\bibnamefont {Ohsumi}},
  \bibinfo {author} {\bibfnamefont {Y.~J.}\ \bibnamefont {Sato}}, \bibinfo
  {author} {\bibfnamefont {R.}~\bibnamefont {Okazaki}},\ and\ \bibinfo {author}
  {\bibfnamefont {D.}~\bibnamefont {Aoki}},\ }\bibfield  {title} {\enquote
  {\bibinfo {title} {{Large transverse thermoelectric effect induced by the
  mixed-dimensionality of Fermi surfaces}},}\ }\href
  {https://doi.org/10.1038/s41467-024-48217-0} {\bibfield  {journal} {\bibinfo
  {journal} {Nat. Commun.}\ }\textbf {\bibinfo {volume} {15}},\ \bibinfo
  {pages} {3907} (\bibinfo {year} {2024})}\BibitemShut {NoStop}%
\bibitem [{\citenamefont {Ohsumi}, \citenamefont {Sato},\ and\ \citenamefont
  {Okazaki}(2024)}]{Ohsumi2024}%
  \BibitemOpen
  \bibfield  {author} {\bibinfo {author} {\bibfnamefont {S.}~\bibnamefont
  {Ohsumi}}, \bibinfo {author} {\bibfnamefont {Y.~J.}\ \bibnamefont {Sato}},\
  and\ \bibinfo {author} {\bibfnamefont {R.}~\bibnamefont {Okazaki}},\
  }\bibfield  {title} {\enquote {\bibinfo {title} {{Transverse Thermoelectric
  Conversion in the Mixed-Dimensional Semimetal WSi$_2$}},}\ }\href
  {https://doi.org/10.1103/PRXEnergy.3.043007} {\bibfield  {journal} {\bibinfo
  {journal} {PRX Energy}\ }\textbf {\bibinfo {volume} {3}},\ \bibinfo {pages}
  {043007} (\bibinfo {year} {2024})}\BibitemShut {NoStop}%
\bibitem [{\citenamefont {Antonova}\ and\ \citenamefont
  {Looman}(2005)}]{Antonova}%
  \BibitemOpen
  \bibfield  {author} {\bibinfo {author} {\bibfnamefont {E.}~\bibnamefont
  {Antonova}}\ and\ \bibinfo {author} {\bibfnamefont {D.}~\bibnamefont
  {Looman}},\ }\bibfield  {title} {\enquote {\bibinfo {title} {{Finite elements
  for thermoelectric device analysis in ANSYS}},}\ }in\ \href
  {https://doi.org/10.1109/ICT.2005.1519922} {\emph {\bibinfo {booktitle} {ICT
  2005. 24th International Conference on Thermoelectrics, 2005.}}}\ (\bibinfo
  {year} {2005})\ pp.\ \bibinfo {pages} {215--218}\BibitemShut {NoStop}%
\bibitem [{\citenamefont {Bose}\ and\ \citenamefont {Hecht}(1992)}]{Bose1992}%
  \BibitemOpen
  \bibfield  {author} {\bibinfo {author} {\bibfnamefont {S.}~\bibnamefont
  {Bose}}\ and\ \bibinfo {author} {\bibfnamefont {R.~J.}\ \bibnamefont
  {Hecht}},\ }\bibfield  {title} {\enquote {\bibinfo {title} {{Thermal
  properties of MoSi$_2$ and SiC whisker-reinforced MoSi$_2$}},}\ }\href
  {https://doi.org/10.1007/BF00540701} {\bibfield  {journal} {\bibinfo
  {journal} {J. Mater. Sci.}\ }\textbf {\bibinfo {volume} {27}},\ \bibinfo
  {pages} {2749--2752} (\bibinfo {year} {1992})}\BibitemShut {NoStop}%
\bibitem [{\citenamefont {Marušić-Paloka}\ and\ \citenamefont
  {Pažanin}(2024)}]{Paloka2024}%
  \BibitemOpen
  \bibfield  {author} {\bibinfo {author} {\bibfnamefont {E.}~\bibnamefont
  {Marušić-Paloka}}\ and\ \bibinfo {author} {\bibfnamefont {I.}~\bibnamefont
  {Pažanin}},\ }\bibfield  {title} {\enquote {\bibinfo {title} {{The Robin
  boundary condition for modelling heat transfer}},}\ }\href
  {https://doi.org/10.1098/rspa.2023.0850} {\bibfield  {journal} {\bibinfo
  {journal} {Proc. R. Soc. A: Math. Phys. Eng. Sci.}\ }\textbf {\bibinfo
  {volume} {480}},\ \bibinfo {pages} {20230850} (\bibinfo {year}
  {2024})}\BibitemShut {NoStop}%
\end{thebibliography}
\end{document}